\documentclass[conference]{IEEEtran}
\usepackage[left=0.75in,right=0.75in,top=0.75in,bottom=0.75in]{geometry}
\usepackage{amsmath,amssymb} \interdisplaylinepenalty=2500
\usepackage{color}
\usepackage{subfigure}
\usepackage{epsfig}
\usepackage{graphicx}
\usepackage{dsfont}
\usepackage{verbatim}
\usepackage{amsthm}
\usepackage{setspace}
\usepackage{mathtools}
\usepackage{enumerate}

\usepackage{tabulary}
\usepackage{booktabs}
\usepackage{bbm}

\usepackage[table]{xcolor}
\usepackage{color, colortbl}
\usepackage{multirow,bigdelim}
\usepackage{amstext}
\usepackage{array}
\usepackage{cite}

\newcount\rowc

\makeatletter
\def\ttabular{%
\hbox\bgroup
\let\\\cr
\def\rulea{\ifnum\rowc=\@ne \hrule height 1.3pt \fi}
\def\ruleb{
\ifnum\rowc=1\hrule height 1.3pt \else
\ifnum\rowc=6\hrule height \heavyrulewidth 
   \else \hrule height \lightrulewidth\fi\fi}
\valign\bgroup
\global\rowc\@ne
\rulea
\hbox to 10em{\strut \hfill##\hfill}%
\ruleb
&&%
\global\advance\rowc\@ne
\hbox to 10em{\strut\hfill##\hfill}%
\ruleb
\cr}
\def\endttabular{%
\crcr\egroup\egroup}

\newtheorem{definition}{Definition}
\newtheorem{theorem}{Theorem}
\newtheorem{lemma}[theorem]{Lemma}

\newenvironment{remark}{\textit{Remark: }}{}

\newif\ifcomments
\commentstrue 

\newcommand{\calB}{\mathcal{B}}

\newcommand{\calE}{\mathcal{E}}

\newcommand{\calH}{\mathcal{H}}

\newcommand{\calN}{\mathcal{N}}

\newcommand{\calT}{\mathcal{T}}
\newcommand{\calU}{\mathcal{U}}

\newcommand{\bfG}{\mathbf{G}}

\newcommand{\bfI}{\mathbf{I}}

\newcommand{\bfS}{\mathbf{S}}

\newcommand{\bfU}{\mathbf{U}}

\newcommand{\bfW}{\mathbf{W}}

\DeclarePairedDelimiterX{\norm}[1]{\lVert}{\rVert}{#1}

\newcommand{\nFile}{K}

\newcommand{\findex}{\kappa}
\newcommand{\nNode}{N}

\newcommand{\queries}{\mathcal{Q}}

\newcommand{\answer}{A}
\newcommand{\secrecy}{S}
\newcommand{\compkprime}{\bar{k'}}
\newcommand{\compk}{\bar{k}}

\newcommand{\Fq}{\mathds{F}_{q}}
\newcommand{\nodeset}{\mathcal{N}}
\newcolumntype{L}{>{$}l<{$}}

\definecolor{Gray}{gray}{0.9}
\definecolor{LightCyan}{rgb}{0.88,1,1}
\definecolor{HoneydewTwo}{rgb}{0.96,0.92,0.88}

\date{}
\begin{document}

\title{\vspace{0.25in} Secure Symmetric Private Information Retrieval from Colluding Databases with Adversaries}
\author{\IEEEauthorblockN{Qiwen~Wang, and~Mikael~Skoglund}
    \IEEEauthorblockA{School of Electrical Engineering, KTH Royal Institute of Technology} \vspace{-1.2em}
    \\Email: \{qiwenw, skoglund\}@kth.se \vspace{-1em}}

\maketitle
\begin{abstract}
The problem of \emph{symmetric private information retrieval (SPIR)} from replicated databases with colluding servers and adversaries is studied. Specifically, the database comprises $K$ files, which are replicatively stored among $N$ servers. A user wants to retrieve one file from the database by communicating with the $N$ servers, without revealing the identity of the desired file to any server. Furthermore, the user shall learn nothing about the other $K-1$ files in the database. Any $T$ out of $N$ servers may collude, that is, they may communicate their interactions with the user to guess the identity of the requested file. 
An adversary in the system can tap in on or even try to corrupt the communication. Three types of adversaries are considered: a Byzantine adversary who can overwrite the transmission of any $B$ servers to the user; a passive eavesdropper who can tap in on the incoming and outgoing transmissions of any $E$ servers; and a combination of both -- an adversary who can tap in on a set of any $E$ nodes, and overwrite the transmission of a set of any $B$ nodes. The problems of SPIR with colluding servers and the three types of adversaries are named T-BSPIR, T-ESPIR and T-BESPIR respectively.
The capacity of the problem is defined as the maximum number of information bits of the desired file retrieved per downloaded bit.
We show that the information-theoretical capacity of the T-BSPIR problem equals $1-\frac{2B+T}{N}$, if the servers share common randomness (unavailable at the user) with amount at least $\frac{2B+T}{N-2B-T}$ times the file size. Otherwise, the capacity equals zero. The information-theoretical capacity of the T-ESPIR problem is proved to equal $1-\frac{\max(T,E)}{N}$, if the servers share common randomness with amount at least $\frac{\max(T,E)}{N-\max(T,E)}$ times the file size. Finally, for the problem of T-BESPIR, the capacity is proved to be $1-\frac{2B+\max(T,E)}{N}$, where the common randomness shared by the servers should be at least $\frac{2B+\max(T,E)}{N-2B-\max(T,E)}$ times the file size. The results resemble those of secure network coding problems with adversaries and eavesdroppers.
\end{abstract}

\section{Introduction}
In the situation where a user wants to retrieve a file from a remotely stored database, the nature of the data might be privacy-sensitive, for example medical records, stock prices {\it etc.},  such that the user does not want to reveal the identity of the data retrieved. This is known as the problem of private information retrieval (PIR). In some cases, the privacy of the database needs also to be preserved. For example, if a user wants to retrieve his/her medical data from a database, it is hoped that the user obtains no information about other users' medical records. This is known as the problem of symmetric private information retrieval (SPIR).

The problem of SPIR was firstly studied in the computer science society.
It is shown that if the database is stored at a single server, the only possible scheme for the user is to download the entire database to guarantee information-theoretic privacy \cite{chor1995private,chor1998private}, which is inefficient in practice. It is further shown that the communication cost can be reduced in sublinear scale by replicating the database at multiple non-colluding servers \cite{chor1998private}.
To further protect the privacy of the database, the problem of SPIR is introduced~\cite{gertner1998protecting}, such that the user obtains no more information regarding the database other than the requested file. 
In~\cite{chor1995private,chor1998private,gertner1998protecting}, the database is modeled as a bit string, and the user wishes to retrieve a single bit.
In these works, the communication cost is measured as the sum of the transmission at the querying phase from user to servers and at the downloading phase from servers to user.

When the file size is significantly large and the target is to minimize the communication cost of only the downloading phase, the metric of the downloading cost is defined as the number of bits downloaded per bit of the retrieved file, and the reciprocal of which is named the \emph{PIR capacity}. 
A series of recent works derive information-theoretic limits of various versions of the PIR problem
\cite{sun2017capacity,sun2016colluding,sun2016SPIR,banawan2016capacity,banawan2017multi,banawan2017capacity,wang2016symmetric} {\it etc}. 
The leading work in the area is by Sun and Jafar\cite{sun2017capacity}, where the authors find the capacity of the PIR problem with replicated databases.
In subsequent works by Sun and Jafar~\cite{sun2016colluding,sun2016SPIR}, the PIR capacity with duplicated databases and colluding servers, and the SPIR capacity with duplicated (non-colluding) databases are derived.
In~\cite{banawan2016capacity,banawan2017multi,banawan2017capacity}, Banawan and Ulukus find the capacity of the PIR problem with coded databases, multi-message PIR with replicated databases, and the PIR problem with colluding and Byzantine databases.
In our previous work~\cite{wang2016symmetric}, we derive the capacity of the SPIR problem with coded databases.


Another series of works focus more on the coding structure of the storage system, and study schemes and information limits for various PIR problems with coded databases~\cite{shah2014one,fazeli2015pir,chan2015private,tajeddine2016private,freij2016private}.
In~\cite{shah2014one}, PIR is achieved by downloading one extra bit other than the desired file, given that the number of storage nodes grows with file size, which can be impractical in some storage systems.
In~\cite{fazeli2015pir}, storage overhead can be reduced by increasing the number of storage nodes.
In~\cite{chan2015private}, tradeoff between storage cost and downloading cost is analyzed. Subsequently in~\cite{tajeddine2016private}, explicit schemes which match the tradeoff in~\cite{chan2015private} are presented.
It is worth noting that in~\cite{banawan2016capacity}, the capacity of PIR for coded database is settled, which improves the results in~\cite{chan2015private,tajeddine2016private}.
Recently in~\cite{freij2016private}, the authors present a framework for PIR from coded database with colluding servers.

In this work, we study the SPIR version of the problem in~\cite{banawan2017capacity}, that is, SPIR from replicated databases with colluding and Byzantine servers. We also study the SPIR problem with a passive eavesdropper, then generalize to the case with an adversary who can both eavesdrop and corrupt the communication.
In analogy to previous works on SPIR~\cite{sun2016SPIR,wang2016symmetric}, in the non-trivial context where the database comprises at least two files, the storage nodes need to share common randomness which is independent from the database and meanwhile unavailable to the user. Furthermore, in the case with an eavesdropper who can tap in on a set of the nodes and is curious about the database, the utility of the shared common randomness is two-fold in the sense that it also protect the database from the eavesdropper.
Briefly speaking, in this work, we study the SPIR problem with replicated databases, where a database with $K$ files are replicated at $N$ servers. Any $T$ out of the $N$ servers may collude, that is, they may share their communication with the user to infer the identity of the requested file. The communication in the system is not secure, that is, there is an adversary who can tap in on or even corrupt the transmissions in the system. We consider three types of adversaries, a Byzantine adversary who can overwrite the transmission of any $B$ servers to the user, named T-BSPIR; a passive eavesdropper who can tap in on the incoming and outgoing transmissions of any $E$ servers, named T-ESPIR; and a combination of both -- an adversary who can tap in on a set of any $E$ nodes, and overwrite the transmission of a set of any $B$ nodes (the two sets may overlap), named T-BESPIR. 
We show that the information-theoretical capacity of the T-BSPIR problem equals $1-\frac{2B+T}{N}$, if the servers share common randomness (unavailable at the user) with amount at least $\frac{T}{N-2B-T}$ times the file size. Otherwise, the capacity equals zero. This is presented in Theorem~\ref{thm:main1}.
The information-theoretical capacity of the T-ESPIR problem is proved to equal $1-\frac{\max(T,E)}{N}$, if the servers share common randomness with amount at least $\frac{\max(T,E)}{N-\max(T,E)}$ times the file size. This is presented in Theorem~\ref{thm:main2}.
Finally in Section~\ref{sec:T-BESPIR}, we show that for the problem of T-BESPIR, the capacity is $1-\frac{2B+\max(T,E)}{N}$, where the common randomness shared by the servers should be at least $\frac{\max(T,E)}{N-2B-\max(T,E)}$ times the file size. The results resemble the capacity of secure network coding with adversaries~\cite{medard2011network}.

\section{Model}
\subsection{Notations}
Let $[m:n]$ denote the set $\{m, m+1, \dots, n\}$ for $m \leq n$. For the sake of brevity, denote the set of random variables $\{ X_m, X_{m+1}, \dots, X_n\}$ by $X_{[m:n]}$ . 
The transpose of matrix $\bfG$ is denoted by $\bfG^{\textrm{T}}$.

\subsection{Problem Description}
\noindent{\bf Database:}
A database comprises $K$ independent files, denoted by $W_1, \dots, W_{K}$, which are replicated at $N$ nodes (servers). Each file consists of $L$ symbols drawn independently and uniformly from the finite field $\Fq$. Therefore, for any $k \in [1:\nFile]$, 
\begin{equation}
H(W_k)=L \log{q} \quad ; \quad H(W_1, \dots, W_{K}) = KL \log{q}. \nonumber
\end{equation}


\noindent{\bf User queries:}
A user wants to retrieve a file $W_{\findex}$ with index $\findex$ from the database, where the desired file index $\findex$ is uniformly distributed among $[1:K]$. 
Let $\calU$ denote a random variable privately generated by the user, which represents the randomness of the query scheme followed by the user. The random variable $\calU$ is generated before the realizations of the messages or the desired file index. Let the realization of the file index $\findex$ be $k$, based on the realization of the desired file index $k$ and the realization of $\calU$, the user generates and sends queries to all nodes, where the query received by node-$n$ is denoted by $Q_{n}^{[k]}$. Let $\queries = [Q_{n}^{[k]}]_{n \in [1:\nNode], k \in [1:\nFile]}$ denote the complete query scheme, namely, the collection of all queries under all cases of desired file index.
We have that $H(\queries | \calU) = 0$.

\noindent{\bf Node common randomness:}
Let random variable $S$ denote the common randomness shared by all nodes, the realization of which is known to all nodes but unavailable to the user. 
The common randomness is utilized to protect database-privacy~\eqref{eqn:data_privacy} below.
For any node $n \in [1:N]$, a random variable $S_n$ is generated from $S$, which is used in the answer scheme followed by node $n$.
Hence, $H(S_1, \dots, S_n | S)=0$.

\noindent{\bf Node answers:}
The nodes generate answers according to the agreed scheme with the user based on the received query $Q_n^{[k]}$, the stored database, and the random variable $S_n$ generated from the common randomness. The answer generated and sent to the user by node $n$ is denoted by $A_{n}^{[k]}$.

\noindent{\bf Adversary:}
Three types of adversaries are considered in this work. The first type is called \emph{Byzantine adversaries}, who can overwrite the answers of a set $\calB$ of at most $B$ nodes, called \emph{corrupted nodes}, pretending to send answers to the user from the corrupted nodes to confuse the user. The nodes that are not corrupted by the adversary are called \emph{authentic nodes}.
The user has no knowledge of the identity of the corrupted nodes. The answers overwritten and sent to the user are denoted by $\tilde{A}_{\calB}^{[k]}$. We assume the Byzantine adversary is omniscient, that is, the adversary can tap in on all transmissions and corrupt the $B$ answers in a worst-case way that confuses the user the most. The model considered in~\cite{banawan2017capacity}, where there are $B$ \emph{Byzantine adversarial nodes} who send arbitrary or worst-case answers to the user, can be considered as a special case where the adversary can only taps on the transmissions of the $B$ nodes chosen to corrupt, {\it i.e.} the adversary has less knowledge.\footnote{For zero-error decodability, the knowledge of the adversary does not affect the result, because the adversary could ``happen" to generate the worst-case corrupted answers without knowing the transmissions in the system, in which case the communication scheme should still prevent the user from decoding the desired file wrong.}

The second type adversary considered is called {\it passive eavesdroppers}, who can tap in on the incoming and outgoing transmissions of $E$ nodes in the system. The eavesdropper is ``nice but curious", in the sense that the goal of the eavesdropper is to obtain some information about the database, without corrupting any transmission. The user has no knowledge of the identity of the nodes tapped on by the eavesdropper.

The third type of adversary considered is a combination of the above two types. The adversary can tap in on the incoming and outgoing communications of any set $\calE$ with $E$ nodes, and can overwrite the answers of any set $\calB$ with $B$ nodes. The two sets may intersect. In this case, the adversary is not omniscient and does not tap in on the nodes that are in $\calB$ but not in $\calE$.

\noindent{\bf T-BSPIR and T-ESPIR:}
Based on the received answers $A_{[1:N]}^{[k]}$ (for the case with Byzantine adversary, we abuse the notation and let $A_{[1:N]}^{[k]} = \{A_{[1:N] \setminus \calB}^{[k]}, \tilde{A}_{\calB}^{[k]} \}$) and the query scheme $\queries$, the user shall be able to decode the requested file $W_{k}$ with zero error. 
Any set of $T$ nodes may collude to guess the requested file index, by communicating their interactions with the user.
Two privacy constraints must be satisfied:
\begin{itemize}
\item \emph{User-privacy:} any $T$ colluding nodes shall not be able to obtain any information regarding the identity of the requested file, {\it i.e.,}
	\begin{equation}
		I(\findex ; Q_{\calT}^{[\findex]}, A_{\calT}^{[\findex]}, W_{[1:K]}, S ) = 0,  \forall \calT \subset [1:\nNode], |\calT|=T. \label{eqn:user_privacy}
	\end{equation}
\item \emph{Database-privacy:} the user shall learn no information regarding other files in the database, that is, defining $W_{\bar{\findex}}= \{ W_1, \dots, W_{\findex-1}, W_{\findex+1}, \dots, W_{\nFile}\}$, 
	\begin{equation}
		I(W_{\bar{\findex}} ; \answer_{[1:\nNode]}^{[\findex]}, \queries, \findex) = 0. \label{eqn:data_privacy}
	\end{equation}
\end{itemize}

For the case with passive eavesdropper and the case with the combination adversary, one more privacy constraint must be satisfied to protect the database from the eavesdropper. For any node set $\calE$ with at most $E$ nodes, and for any $k \in [1:K]$:
\begin{equation}
	I( W_{[1:K]} ; Q_{\calE}^{[k]}, A_{\calE}^{[k]}) = 0. \label{eqn:privacy_E}
\end{equation}

We use the same definition as in~\cite{wang2016symmetric} for rate and capacity of T-BSPIR, T-ESPIR and T-BESPIR schemes. (We state only the definitions in terms of T-BSPIR.)
\begin{definition} 
The rate of a T-BSPIR scheme is the number of information bits of the requested file retrieved per downloaded answer bit. By symmetry among all files, for any $k \in [1:K]$,
\begin{equation}
R_{\textrm{T-BSPIR}} \triangleq \frac{H(W_{k})}{\sum_{n=1}^{N} H(A_n^{[k]})}. \nonumber
\end{equation}
The capacity $C_{\textrm{T-BSPIR}}$ is the supremum of $R_{\textrm{T-BSPIR}}$ over all T-BSPIR schemes.
\end{definition}

\begin{definition} 
The secrecy rate is the amount of common randomness shared by the storage nodes relative to the file size, that is
\begin{equation}
\rho_{\textrm{T-BSPIR}} \triangleq \frac{H(S)}{H(W_{k})}. \nonumber
\end{equation}
\end{definition}

\section{Main Result} \label{sec:main}
\subsection{T-BSPIR}
When there is only one file in the database, {\it i.e.} $K=1$, database-privacy is guaranteed automatically, because there is no other file to protect from the user in the database. Therefore, the T-BSPIR problem reduces to T-BPIR problem, and from~\cite{banawan2017capacity}, the capacity is $1-\frac{2B}{N}$ if $N>2B+T$. In fact, when $K=1$, user-privacy is also trivial, since there is only one file that the user can request for. That is the reason the parameter $T$ is not in the capacity $1-\frac{2B}{N}$. Therefore, the condition can be relaxed to that if $N \geq 2B+1$, the capacity of T-BSPIR when $K=1$ is $1-\frac{2B}{N}$. If $N \leq 2B$, the user cannot successfully retrieve the file regardless of how much information downloaded, {\it i.e.} the capacity is $0$.
When $K \geq 2$, T-BSPIR is non-trivial and our main result is summarized below.

\begin{theorem}
For symmetric private information retrieval from a database with $K \geq 2$ files which are replicated at $N$ nodes, where any $T$ nodes may collude and a Byzantine adversary can corrupt the answers of any $B$ nodes, if $N > 2B+T$, the capacity is 
\begin{equation}
C_{\textrm{T-BSPIR}} = 
\begin{cases}
1-\frac{2B+T}{N}, & \text{if } \rho_{\textrm{T-BSPIR}} \geq \frac{T}{N-2B-T}\\
0, & \text{otherwise}
\end{cases}
. \nonumber
\end{equation} 
\label{thm:main1}
\end{theorem} 

\begin{remark}
In~\cite{banawan2017capacity}, the authors show that the T-BPIR capacity is $\frac{N-2B}{N} \cdot \frac{1-\frac{T}{N-2B}}{1-(\frac{T}{N-2B})^K}$.
It can be observed that as the number of files $K$ tends to infinity, 
their T-BPIR capacity approaches our T-BSPIR capacity.
The intuition is that, when the number of files increases, the penalty in the downloading rate to protect database-privacy decays.
When there are asymptotically infinitely many files, the information rate the user can learn about the database from finite downloaded symbols vanishes.
\end{remark}

\subsection{T-ESPIR}
When there is only one file in the database, the database-privacy and user-privacy become trivial. The only privacy constraint needed to be guaranteed is that the eavesdropper learns no information of the database~\eqref{eqn:privacy_E}. It can be easily checked that the capacity equals $1-\frac{E}{N}$. When $K \geq 2$, the capacity of T-ESPIR is summarized below.
\begin{theorem}
For symmetric private information retrieval from a database with $K \geq 2$ files which are replicated at $N$ nodes, where any $T$ nodes may collude and an eavesdropper can tapped on the communication of any $E$ nodes, the capacity is 
\begin{equation}
C_{\textrm{T-ESPIR}} = 
\begin{cases}
1-\frac{\max{(T,E)}}{N}, & \text{if } \rho_{\textrm{T-ESPIR}} \geq \frac{\max{(T,E)}}{N-\max{(T,E)}}\\
0, & \text{otherwise}
\end{cases}
. \nonumber
\end{equation} 
\label{thm:main2}
\end{theorem}

\section{T-BSPIR}
\subsection{Achievability} \label{sec:achieve_B}
In this section, we present a general scheme which achieves the maximum T-BSPIR rate $1 - \frac{2B+T}{N}$ when the secrecy rate is $\frac{T}{N-2B-T}$.
The main concepts of the construction are:
\begin{itemize}
\item The queries received by any set of $T$ nodes are mutually independent, and are independent of the desired file index $k$. This is achieved by expanding $T$ independent query vectors with an $(N,T)$-MDS code.
\item Because the answers received from any $B$ nodes might be erroneous, the $N$ answers are formalized in a form of $(N, N-2B)$-MDS code, such that the user can correct up to $B$ errors.
\end{itemize}

Assume each file comprises $L = N-2B-T$ symbols from a large enough field $\Fq$.\footnote{The field size should be large enough such that the MDS codes used in the construction exist.} Let the vector $\bfW = (w_{1}^{[1]}, \dots, w_{N-2B-T}^{[1]}, \dots, w_{1}^{[K]}, \dots, w_{N-2B-T}^{[K]})$ represent the database, which is stored at each server.
The user wants to retrieve $W_k = (w_{1}^{[k]}, \dots, w_{N-2B-T}^{[k]})$ privately.

The user generates the queries following the steps below:
\emph{Step 1:} Generates $T$ independent uniformly random vectors $U_1, \dots, U_T$ of length $K(N-2B-T)$ over $\Fq$. Let the $K(N-2B-T) \times T$ matrix $\bfU$ denote $[U_1, \dots, U_T]$.

\emph{Step 2:} Let $e_i^{[k]}$ denote the length-$(K(N-2B-T))$ unit vector where only the $\big( (k-1)(N-2B-T)+i \big)$th entry is $1$ and all the other entries are $0$'s. The purpose of $e_i^{[k]}$ is to retrieve the $i$th entry of $W_k$. Let the  $K(N-2B-T) \times (N-2B-T)$ matrix $\mathbf{e}$ denote $[e_1^{[k]}, e_2^{[k]}, \dots, e_{N-2B-T}^{[k]}]$.

\emph{Step 3:} Let $\{ \lambda_1, \dots, \lambda_N \}$ be $N$ distinct nonzero elements from $\Fq$. Let $\bfG_{\bfU}$ be the generating matrix of an $(N,T)$-generalized-Reed-Solomon (GRS) code with code locators  $\{ \lambda_1, \dots, \lambda_N \}$ and column multipliers all be $1$. That is, 
\begin{equation}
{
\bfG_{\bfU} = 
\begin{bmatrix}
1 & 1 & \dots & 1\\
\lambda_1 & \lambda_2 & \dots & \lambda_N         \\
\vdots  & \vdots   & \ddots & \vdots   \\
\lambda_1^{T-1} & \lambda_2^{T-1} &  \dots & \lambda_N^{T-1}   
\end{bmatrix}
}
. 
\end{equation}
Let $\bfG_{\mathbf{e}}$ be the generating matrix of an $(N,N-2B-T)$-GRS code with code locators  $\{ \lambda_1, \dots, \lambda_N \}$ and column multipliers $\{  \lambda_1^{T-1}, \dots, \lambda_N^{T-1}\} $. That is, 
\begin{align}
\bfG_{\mathbf{e}} 
& = 
\begin{bmatrix}
1 & 1 & \dots & 1\\
\lambda_1 & \lambda_2 & \dots & \lambda_N         \\
\vdots  & \vdots   & \ddots & \vdots   \\
\lambda_1^{N-2B-T-1} & \lambda_2^{N-2B-T-1} &  \dots & \lambda_N^{N-2B-T-1}   
\end{bmatrix}  \\
& \cdot diag( \lambda_1^{T-1}, \lambda_2^{T-1}, \dots, \lambda_N^{T-1}) \\ 
& = 
\begin{bmatrix}
\lambda_1^{T} & \lambda_2^{T} &  \dots & \lambda_N^{T}\\
\lambda_1^{T+1} & \lambda_2^{T+1} & \dots & \lambda_N^{T+1}         \\
\vdots  & \vdots   & \ddots & \vdots   \\
\lambda_1^{N-2B-1} & \lambda_2^{N-2B-1} &  \dots & \lambda_N^{N-2B-1}   
\end{bmatrix}
. 
\end{align}

\emph{Step 4:} Generate the $N$ query vectors $Q_1^{[k]}, \dots, Q_n^{[k]}$ by
\begin{align}
[ Q_1^{[k]}, \dots, Q_n^{[k]} ]
& = \bfU \bfG_{\bfU} + \mathbf{e} \bfG_{\mathbf{e}} \\
& = [\bfU , \mathbf{e}] \cdot
\begin{bmatrix}
\bfG_{\bfU} \\
\bfG_{\mathbf{e}} \label{eqn:query_vectors}
\end{bmatrix}
\end{align}

The user sends the query vectors generated from equation~\eqref{eqn:query_vectors} to the servers. 

All the servers share $T$ symbols $S_1, \dots, S_T$ that are uniformly and independently chosen from $\Fq$, which are unavailable to the user. 
The servers generate their answers by taking the inner product of the query vectors they receive and the stored data vector, then add on a linear combination of $S_1, \dots, S_T$. Specifically, 
\begin{equation}
A_n^{[k]} = \langle Q_n^{[k]}, \bfW \rangle + \sum_{j=1}^{T} \lambda_n^{j-1} S_j. \label{eqn:answer_n}
\end{equation}
There are at most $B$ servers corrupted by the Byzantine adversary, who generate arbitrary (or even malicious) answers $\tilde{A}_n^{[k]}$ to confuse the user. Assume the Byzantine adversary generates answers of the same size as the authentic servers, {\it i.e.} the size the user expects to receive, otherwise the user can easily identify the erroneous answers.

To see that the user can decode $W_k$ successfully, firstly we look at the $N$ correct answers. Denote $X_j = \langle U_j, \bfW \rangle +S_j$, where $j = 1, \dots, T$. From~\eqref{eqn:query_vectors} and~\eqref{eqn:answer_n}, $A_n^{[k]} = X_1 + \lambda_n X_2 + \cdots + \lambda_n^{T-1} X_T + \lambda_n^{T} w_1^{[k]} + \cdots + \lambda_n^{N-2B-1} w_{N-2B-T}^{[k]}$. Hence,
\begin{equation}
[ A_1^{[k]}, \dots, A_n^{[k]} ] = [X_1, \dots, X_T, w_1^{[k]}, \dots, w_{N-2B-T}^{[k]}] \cdot \bfG,
\end{equation}
where
\begin{align}
\bfG  & =  [\bfG_{\bfU}, \bfG_{\mathbf{e}}]^{\textrm{T}} \nonumber \\
& = 
\begin{bmatrix}
1 & 1 & \dots & 1\\
\lambda_1 & \lambda_2 & \dots & \lambda_N         \\
\vdots  & \vdots   & \ddots & \vdots   \\
\lambda_1^{N-2B-1} & \lambda_2^{N-2B-1} &  \dots & \lambda_N^{N-2B-1}   
\end{bmatrix}.
\end{align}
It can be observed that $\bfG$ is the generating matrix of an $(N,N-2B)$-GRS code with code locators  $\{ \lambda_1, \dots, \lambda_N \}$ and column multipliers all be $1$. Therefore, when at most $B$ symbols out of $[ A_1^{[k]}, \dots, A_n^{[k]} ]$ are wrong, the user can still successfully decode $[X_1, \dots, X_T, w_1^{[k]}, \dots, w_{N-2B-T}^{[k]}]$, which include all symbols of $W_k$.

It is obvious that database-privacy is guaranteed. Because besides $W_k$, the user solves $T$ symbols $X_j = \langle U_j, \bfW \rangle +S_j$, $j = 1, \dots, T$. Because $S_1, \dots, S_T$ are independent uniform symbols drawn from $\Fq$, the user can obtain no information about the linear combinations of the database from the $X_j$'s.

To see that user-privacy is guaranteed, from~\eqref{eqn:query_vectors}, because $U_1, \dots, U_T$ are $T$ independent random vectors and $\bfG_{\bfU}$ is the generating matrix of an $(N,T)$-GRS code, any $T$ column vectors of $\bfU \bfG_{\bfU}$ are still independent uniform random vectors. Hence, by adding deterministic column vectors of $\mathbf{e} \bfG_{\mathbf{e}}$, any $T$ query vectors are still independent uniform random vectors, and are independent from the desired file index. Therefore, any $T$ colluding nodes cannot infer the desired file index.

To conclude, the rate achieved by this scheme is $\frac{N-2B-T}{N} = 1- \frac{2B+T}{N}$ with secrecy rate $\frac{T}{N-2B-T}$, which matches the capacity.

\subsection{Converse} \label{sec:converse_B}
In this section, we prove the converse part of Theorem~\ref{thm:main1}. Lemmas~\ref{thm:lemma_noQ}-\ref{thm:lemma_noWk} below are the versions with colluding servers and replicated databases of Lemmas 2-4 in~\cite{wang2016symmetric} (and Lemmas 1-2 in~\cite{sun2016SPIR}). Hence we state the lemmas with sketch proofs. 
For any set of nodes that are not corrupted by the adversary, given their received queries, the answers generated by these nodes do not depend on other queries. Because besides the received queries, the answers depend on the database and the shared common randomness, which are independent with other queries.
Lemma~\ref{thm:lemma_noQ} below states that this also holds if conditioned on the requested file.

\begin{lemma} \label{thm:lemma_noQ}
For any set of nodes $\nodeset \subset [1:\nNode]$ that are not corrupted by the adversary, 
	\begin{equation}
		H(A_{\nodeset}^{[k]} | \queries, W_k, Q_{\nodeset}^{[k]}) = H(A_{\nodeset}^{[k]} | W_k, Q_{\nodeset}^{[k]}). \nonumber
	\end{equation}	
\end{lemma}
\noindent {\it Proof:}
We first show that $I(A_{\nodeset}^{[k]}; \queries | W_k, Q_{\nodeset}^{[k]}) \leq  0$, as follows
\begin{align*}
 I(A_{\nodeset}^{[k]}; \queries | W_k, Q_{\nodeset}^{[k]}) 
& \leq I(A_{\nodeset}^{[k]}, W_{[1:K]}, \secrecy ; \queries | W_k, Q_{\nodeset}^{[k]}) \\
& \stackrel{(a)}{=} I(W_{[1:K]}, \secrecy ; \queries | W_k, Q_{\nodeset}^{[k]}) \\
& \leq I(W_{[1:K]}, \secrecy ; \queries) = 0,
\end{align*}
where $(a)$ holds because the answers are deterministic functions of the database, common randomness, and the queries.
In the last step, $I(W_{[1:K]}, \secrecy ; \queries) = 0$ holds because the queries do not depend on the database and  common randomness.

On the other hand, it is immediate that $I(A_{\nodeset}^{[k]}; \queries | W_k, Q_{\nodeset}^{[k]}) \geq  0$. Therefore, $H(A_{\nodeset}^{[k]}| W_k, Q_{\nodeset}^{[k]}) =  H(A_{\nodeset}^{[k]} | \queries, W_k, Q_{\nodeset}^{[k]})$.
\hfill $\Box$

\begin{lemma} \label{thm:lemma_kkprime}
For any set of nodes $\calT \subset [1:N]$ with size $|\calT| = T$ that are not corrupted by the adversary,
	\begin{equation} 
		H(A_{\calT}^{[k]} |  Q_{\calT}^{[k]}) = H(A_{\calT}^{[k']}|  Q_{\calT}^{[k']}), \label{eqn:lemma_kkprime1}
	\end{equation}
	\begin{equation} 
		H(A_{\calT}^{[k]} | W_k, Q_{\calT}^{[k]}) = H(A_{\calT}^{[k']} | W_k, Q_{\calT}^{[k']}). \label{eqn:lemma_kkprime2}
	\end{equation}	
\end{lemma}
\noindent {\it Proof:}
The proof is similar as that of Lemma 1 in~\cite{sun2016SPIR}. We omit the detailed proof here. The key idea is that since any $T$ nodes may collude, the statistical distribution of the queries and answers of any $T$ nodes shall be the same regardless of the requested file index, even if the nodes condition on a part of the database, for example $W_k$. Otherwise, the $T$ nodes can differentiate between the cases where $W_k$ is requested and $W_{k'}$ is requested.
\hfill $\Box$

\begin{lemma} \label{thm:lemma_noWk}
For any set of nodes $\calT \subset [1:N]$ with size $|\calT| = T$ that are not corrupted by the adversary,
	\begin{equation}
		H(A_{\calT}^{[k]} | W_k, Q_{\calT}^{[k]}) = H(A_{\calT}^{[k']} |  Q_{\calT}^{[k']}).\nonumber
	\end{equation}	
\end{lemma}
\noindent {\it Proof:}
By database-privacy~\eqref{eqn:data_privacy}, $ I(W_{\bar{k}} ; A_{[1:N]}^{[k']}, \queries) = 0$.
For any $k \neq k'$, because $W_k \in W_{\compkprime}$, we have
\begin{align*}
0
& = I(W_{k} ; A_{\calT}^{[k']}, Q_{\calT}^{[k']}) \\
& = I(W_{k} ; A_{\calT}^{[k']}| Q_{\calT}^{[k']}) +  I(W_{k} ; Q_{\calT}^{[k']}) \\
& \stackrel{(a)}{=}   I(W_{k} ; A_{\calT}^{[k']}| Q_{\calT}^{[k']}) \\
& = H(A_{\calT}^{[k']}| Q_{\calT}^{[k']}) - H(A_{\calT}^{[k']}| W_k, Q_{\calT}^{[k']}) \\
& \stackrel{(b)}{=} H(A_{\calT}^{[k']}| Q_{\calT}^{[k']}) - H(A_{\calT}^{[k]}| W_k, Q_{\calT}^{[k]}) ,
\end{align*}
where equality $(a)$ holds because $W_k$ is independent of the queries, and equality $(b)$ follows by~\eqref{eqn:lemma_kkprime2}.
\hfill $\Box$

Lemma~\ref{thm:lemma_honestdecodability} below states that the user should be able to decode the desired file from any $N-2B$ authentic nodes. This is similar as Lemma 4 in~\cite{banawan2017capacity}, developed from the cut-set bound in the network coding problem~\cite{jaggi2007resilient,kosut2014polytope}, and the distributed storage problem~\cite{pawar2011securing}. 
The difference between our Lemma~\ref{thm:lemma_honestdecodability} and Lemma 4 in~\cite{banawan2017capacity} is that instead of arguing that the answers from any $N-2B$ authentic nodes must be unique for every realization of the database, we argue that it only needs to hold for any realization of the requested file $W_k$. For different realizations of the database that differ on files other than $W_k$, the interference may still be the same hence the user can successfully decode.
We reprise the proof of Lemma 4 in~\cite{banawan2017capacity} with slight modification for the proof of Lemma~\ref{thm:lemma_honestdecodability} below.

\begin{lemma} \label{thm:lemma_honestdecodability}
For any set of authentic nodes $\calH \in [1:N]$ where $|\calH| = N-2B$, for correctly decoding $W_k$, the answers $A_{\calH}^{[k]}$ are unique for every realization of $W_k$. That is, there cannot exist two realizations of the $k$th file, $W_k \neq \tilde{W}_k$, such that $A_{\calH}^{[k]} (W_k) = A_{\calH}^{[k]} (\tilde{W}_k)$.
Consequently, $H(W_k|A_{\calH}^{[k]} ,\queries)=0$.
\end{lemma}
\noindent {\it Proof:}
Divide the nodes $[1:N] \setminus \calH$ into two size-$B$ sets, denoted by $\calB_1$ and $\calB_2$. The scheme shall allow the user to correctly decode $W_k$ if any $B$ nodes in $[1:N] \setminus \calH$ are corrupted by the Byzantine adversary, with any corrupted answers. Consider the following two cases:

\begin{itemize}
\item Case 1: The true realization of the $k$th file is $W_k$. The user downloads $A_{\calH}^{[k]} (W_k)$ from the authentic nodes in $\calH$. The nodes in $\calB_1$ are also authentic, who generates the answers $A_{\calB_1}^{[k]} (W_k)$. The nodes in $\calB_2$ are the corrupted nodes, the answers from which overwritten by the adversary ``happened" to be generated with the agreed scheme but by replacing the $k$th file with $\tilde{W}_k$, denoted by $\tilde{A}_{\calB_2}^{[k]} =  A_{\calB_2}^{[k]}(\tilde{W}_k)$. 

\item Case 2: The true realization of the $k$th file is $\tilde{W}_k$. The user downloads $A_{\calH}^{[k]} (\tilde{W}_k)$ from the authentic nodes in $\calH$. The nodes in $\calB_2$ are also authentic, who generates the answers $A_{\calB_2}^{[k]} (\tilde{W}_k)$. The nodes in $\calB_1$ are corrupted, the answers from which overwritten by the adversary ``happened" to be generated with the agreed scheme but by replacing the $k$th file with $W_k$, hence generating $\tilde{A}_{\calB_1}^{[k]} = A_{\calB_1}^{[k]} (W_k)$.
\end{itemize}

If $A_{\calH}^{[k]} (W_k) = A_{\calH}^{[k]} (\tilde{W}_k)$, under both cases, the user downloads the same set of answers from all nodes, {\it i.e.,} $\left( A_{\calH}^{[k]} (W_k) = A_{\calH}^{[k]} (\tilde{W}_k), A_{\calB_1}^{[k]} (W_k), A_{\calB_2}^{[k]} (\tilde{W}_k)\right)$. Hence, the user cannot successfully decode whether the $k$th file is $W_k$ or $\tilde{W}_k$. 

In conclusion, for any different realization of $W_k$, the answers from $\calH$ differs. In other words, the user should be able to successfully decode the desired file from the $N-2B$ authentic nodes, {\it. i.e.,} $H(W_k|A_{\calH}^{[k]} ,\queries)=0$.
\hfill $\Box$

\subsubsection{The proof for $R \leq C_{\textrm{T-BSPIR}}$ } \label{section:pfThm1_1}
By Lemma~\ref{thm:lemma_honestdecodability}, let $\calH$ be a set of $N-2B$ honest nodes, $N-2B \geq T$,
\begin{align*}
H(W_k) 
& = H(W_k) - H(W_k | A_{\calH}^{[k]} ,\queries) \\
& = I(W_k; A_{\calH}^{[k]} |\queries) \\
& = H(A_{\calH}^{[k]} |\queries) - H(A_{\calH}^{[k]} |W_k, \queries) \\
& \stackrel{(a)}{\leq} H(A_{\calH}^{[k]} |\queries) - H(A_{\calT}^{[k]} |W_k, \queries) \\
& \stackrel{(b)}{=} H(A_{\calH}^{[k]} |\queries) - H(A_{\calT}^{[k]} |W_k, Q_{\calT}^{[k]}) \\
&  \stackrel{(c)}{=} H(A_{\calH}^{[k]} |\queries) - H(A_{\calT}^{[k']} |Q_{\calT}^{[k']}) \\
&  \stackrel{(d)}{=} H(A_{\calH}^{[k]} |\queries) - H(A_{\calT}^{[k]} |Q_{\calT}^{[k]}) \\
&  \leq H(A_{\calH}^{[k]} |\queries) - H(A_{\calT}^{[k]} |\queries) 
\end{align*}

In step (a), $\calT$ can be any set of $T$ nodes in $\calH$. Step (b) holds by Lemma~\ref{thm:lemma_noQ}. Steps (c) and (d) follow by Lemma~\ref{thm:lemma_noWk} and Lemma~\ref{thm:lemma_kkprime} respectively.

Averaging over all $\calT$ with size $T$ from $\calH$, we have that
\begin{equation}
H(W_k) \leq H(A_{\calH}^{[k]} |\queries) - \frac{1}{{N-2B \choose T}}\sum_{\substack{ \calT \in \calH \\ |\calT|=T}} H(A_{\calT}^{[k]} | \queries). \nonumber
\end{equation}

By Han's inequality~\cite{cover2012elements}, $$\frac{1}{{N-2B \choose T}}\sum_{\substack{ \calT \in \calH \\ |\calT|=T}} H(A_{\calT}^{[k]} | \queries) \geq \frac{T}{N-2B} H(A_{\calH}^{[k]} | \queries).$$

Hence, $H(W_k) \leq \frac{N-2B-T}{N-2B} H(A_{\calH}^{[k]} | \queries) \leq \frac{N-2B-T}{N-2B} (N-2B) H(A_{h_1}^{[k]} | \queries) \leq (N-2B-T) H(A_{h_1}^{[k]}|\queries)$, where $h_1 \in \calH$ is an honest node.

Assume that the corrupted nodes send the same amount of information bits to the user, otherwise the user can easily identify the corrupted nodes. Hence,
$R_{\textrm{T-BSPIR}}  = \frac{H(W_k)}{\sum_{n=1}^{\nNode} H(A_n^{[k]})} = \frac{H(W_k)}{N \cdot H(A_{h_1}^{[k]})} \leq 1 - \frac{2B+T}{N}$.

\subsubsection{The proof for $\rho_{\textrm{T-BSPIR}} \geq \frac{T}{N-2B-T}$}
By database-privacy,
\begin{align*}
0 
& =  I(W_{\compk} ; A_{\calH}^{[k]} | \queries) \\
& = H(W_{\compk} | \queries) - H(W_{\compk} | A_{\calH}^{[k]}  , \queries) \\
& \stackrel{(a)}{=} H(W_{\compk} | \queries, W_k) - H(W_{\compk} | A_{\calH}^{[k]}  , \queries, W_k) \\
& = I(W_{\compk} ; A_{\calH}^{[k]}  | \queries, W_k) \\
& \stackrel{(b)}{\geq} I(W_{\compk} ; A_{\calT}^{[k]} | \queries, W_k) \\
& \stackrel{(c)}{=}  H(A_{\calT}^{[k]} | \queries,  W_k)  -   H(A_{\calT}^{[k]} | \queries, W_{[1:K]})  \\
& \qquad +  H(A_{\calT}^{[k]} | \queries, W_{[1:K]}, S) \\
& = H(A_{\calT}^{[k]} | \queries, W_k) - I(S ; A_{\calT}^{[k]} | \queries, W_{[1:K]}) \\
& \geq  H(A_{\calT}^{[k]} | \queries) - H(S),
\end{align*}
where step (a) follows from Lemma~\ref{thm:lemma_honestdecodability} that the user should be able to decode $W_k$ from $A_{\calH}^{[k]}$. In step (b), $\calT$ can be any set of $T$ nodes in $\calH$. Step (c) holds because the authentic answers are deterministic functions of the queries, the database, and the common randomness.

Averaging over all $\calT \subset \calH$, and from the proof in Section~\ref{section:pfThm1_1} above,
\begin{align*}
H(S)
& \geq \frac{1}{{N-2B \choose T}}\sum_{\substack{ \calT \subset calH \\ |\calT|=T}} H(A_{\calT}^{[k]} | \queries) \\
& \geq \frac{T}{N-2B} H(A_{\calH}^{[k]} | \queries) \\
& \geq \frac{T}{N -2B-T} H(W_k).
\end{align*}
Hence, $\rho_{\textrm{T-BSPIR}}  = \frac{H(S)}{H(W_k)} \geq \frac{T}{N-2B-T}$.

\section{T-ESPIR}
\subsection{Achievability} \label{sec:achieve_E}
Assume each file comprises $L = N-\max(T,E)$ symbols from a large enough field $\Fq$. Let the vector $\bfW = (w_{1}^{[1]}, \dots, w_{N-\max(T,E)}^{[1]}, \dots, w_{1}^{[K]}, \dots, w_{N-\max(T,E)}^{[K]})$ represent the database, which is stored at each server.
The user wants to retrieve $W_k = (w_{1}^{[k]}, \dots, w_{N-\max(T,E)}^{[k]})$ privately.

The queries are generated in the following way. 
The user firstly generate $\max(T,E)$ independent uniformly random vectors $U_1, \dots, U_{\max(T,E)}$ of length $K(N-\max(T,E))$ over $\Fq$. The user choose an $(N, \max(T,E))$-GRS code with generating matrix $\bfG_{(N,\max{(T,E)})}$.
Let $e_i^{[k]}$ denote the length-$(K(N-\max(T,E)))$ unit vector where only the $\big( (k-1)(N-\max(T,E))+i \big)$th entry is $1$ and all the other entries are $0$'s. Again, the purpose of $e_i^{[k]}$ is to retrieve the $i$th entry of $W_k$. 
The query vectors are generated by
\begin{align} 
[Q_1^{[k]}, \dots, Q_N^{[k]}] & = [U_1, \dots, U_{\max(T,E)}] \cdot \bfG_{(N,\max{(T,E)})} \nonumber \\
& \qquad + [0, \dots, 0, e_1^{[k]}, \dots, e_{N-max(T,E)}^{[k]}]. \label{eqn:queries_E}
\end{align}

The nodes share $\max(T,E)$ symbols $(S_1, \dots, S_{\max(T,E)}) = \bfS$, called common randomness, that are uniformly and independently chosen from $\Fq$. The common randomness is unavailable to the user and the eavesdropper. The servers generate their answers by taking the inner product of the query vector and the stored data vector, then add on a linear combination of the common randomness in the following way,
\begin{equation} \label{eqn:answern_E}
A_n^{[k]} = \langle Q_n^{[k]}, \bfW \rangle + \langle \bfG_{(N,\max{(T,E)})} (n), \bfS \rangle ,
\end{equation}
where $\bfG_{(N,\max{(T,E)})} (n)$ denotes the $n$th column of matrix $\bfG_{(N,\max{(T,E)})}$.
Let $X_j = \langle U_j, \bfW \rangle +S_j$, where $j = 1, \dots, \max(T,E)$, the answers received by the user are
\begin{align} 
[A_1^{[k]}, \! \dots \! , A_N^{[k]}] & \! = \! [X_1, \! \dots \!, X_{\max(T,E)}, w_1^{[k]}, \! \dots \! , w_{N-\max(T,E)}^{[k]}] \nonumber \\
& \quad \cdot 
\begin{bmatrix}
  \bfG_{(N,\max{(T,E)})} \\
\mathbf{0} \; \; \; \; \; \bfI
\end{bmatrix}, \label{eqn:answers_E}
\end{align}
where we omit the dimension of the zero matrix $\mathbf{0}$ and the identity matrix $\bfI$ because there is no ambiguity.
Because $\bfG_{(N,\max{(T,E)})}$ is the generating matrix of an $(N,\max(T,E))$-GRS code, the matrix $\begin{bmatrix}
  \bfG_{(N,\max{(T,E)})} \\
\mathbf{0} \; \; \; \; \; \bfI
\end{bmatrix}$
is invertible. Therefore, the user can solve $[X_1, \dots, X_{\max(T,E)}, w_1^{[k]}, \dots, w_{N-\max(T,E)}^{[k]}]$, hence obtain $W_k$.

To see that database-privacy is guaranteed, besides the symbols of $W_k$, the user solves $X_1, \dots, X_{\max(T,E)}$, where $X_j = \langle U_j, \bfW \rangle +S_j$. Because $S_1, \dots, S_{\max(T,E)}$ are independent uniform symbols drawn from $\Fq$, the user can obtain no information about the database.
User-privacy is also guaranteed, because from equation~\eqref{eqn:queries_E}, every $\max(T,E)$ query vectors are independently and uniformly distributed. Hence every $T$ nodes see independent and uniformly distributed query vectors, no matter which file the user requests.
To see that the eavesdropper learns no information about the database, the eavesdropper taps on the queries and answers of $E$ nodes.
By the MDS property of GRS codes, any $E$ columns of $\bfG_{(N,\max{(T,E)})}$ are linearly independent.  From equation~\eqref{eqn:answern_E}, any $E$ answers are protected by independent linear combinations of $S_1, \dots, S_{\max(T,E)}$. That is, for any $E$ nodes $n_1, \dots, n_E$, $ \langle \bfG_{(N,\max{(T,E)})} (n_i), \bfS \rangle $'s are statistically independent and uniformly distributed. Hence, from any $E$ query and answer pairs, the eavesdropper obtains no information about the database, {\it i.e.}~\eqref{eqn:privacy_E} is satisfied.

\subsection{Converse} \label{sec:converse_E}
In this section, we prove the converse part of Theorem~\ref{thm:main2}. We also use Lemmas~\ref{thm:lemma_noQ}-\ref{thm:lemma_noWk} in Section~\ref{sec:converse_B} for the proofs below.

\subsubsection{The proof for $R \leq C_{\textrm{T-ESPIR}}$ } \label{sec:pfThm2_1}
For any file $W_k$, $k \in [1:K]$, and any set of nodes $\nodeset \in [1:N]$ with size $|\nodeset|=\max{(T,E)}$,
\begin{align*}
H(W_k) 
& = H(W_k|\queries) - H(W_k|A_{[1:\nNode]}^{[k]}, \queries) \\
& = H(A_{[1:\nNode]}^{[k]} | \queries) - H(A_{[1:\nNode]}^{[k]} | W_k, \queries) \\
& \leq H(A_{[1:\nNode]}^{[k]} | \queries) - H(A_{\nodeset}^{[k]} | W_k, \queries, Q_{\nodeset}^{[k]}) \\
& \stackrel{(a)}{=} H(A_{[1:\nNode]}^{[k]} | \queries) - H(A_{\nodeset}^{[k]} | W_k, Q_{\nodeset}^{[k]}) \\
& \stackrel{(b)}{=} H(A_{[1:\nNode]}^{[k]} | \queries) - H(A_{\nodeset}^{[k]} |  Q_{\nodeset}^{[k]}) \\
& \leq H(A_{[1:\nNode]}^{[k]} | \queries) - H(A_{\nodeset}^{[k]} | \queries),
\end{align*}
where $(a)$ holds because given the queries $Q_{\nodeset}^{[k]}$, the answers of $\nodeset$ do not depend on other queries.
If $\max(T,E)=T$, by Lemma~\ref{thm:lemma_noWk} and Lemma~\ref{thm:lemma_kkprime}, we have that $(b)$ holds; if $\max(T,E)=E$, from equation~\eqref{eqn:privacy_E}, $I(W_k; A_{\nodeset}^{[k]} , Q_{\nodeset}^{[k]}) = 0$, hence $(b)$ also holds.

Averaging over all $\nodeset$ with size $\max(T,E)$, we have that
\begin{equation}
H(W_k) \leq H(A_{[1:\nNode]}^{[k]} | \queries) - \frac{1}{{N \choose \max(T,E)}}\sum_{\substack{ \nodeset \in [1:N] \\ |\nodeset|=\max(T,E)}} H(A_{\nodeset}^{[k]} | \queries). \nonumber
\end{equation}

By Han's inequality~\cite{cover2012elements}, $$\frac{1}{{N \choose \max(T,E)}}\sum_{\substack{ \nodeset \in [1:N] \\ |\nodeset|=\max(T,E)}} H(A_{\nodeset}^{[k]} | \queries) \geq \frac{\max(T,E)}{N} H(A_{[1:\nNode]}^{[k]} | \queries).$$

Therefore, 
$R_{\textrm{T-ESPIR}} = \frac{H(W_k)}{\sum_{n=1}^{\nNode} H(A_n^{[k]})} \leq \frac{H(W_k)}{H(A_{[1:\nNode]}^{[k]} | \queries)} \leq 1 - \frac{\max(T,E)}{N}$.

\subsubsection{The proof for $\rho_{\textrm{T-ESPIR}} \geq \frac{\max{(T,E)}}{N-\max{(T,E)}}$}
For any set of nodes $\nodeset \in [1:N]$ with size $|\nodeset|=\max{(T,E)}$, from database-privacy~\eqref{eqn:data_privacy},
\begin{align*}
0 
& =  I(W_{\compk} ; A_{[1:\nNode]}^{[k]} | \queries) \\
& = H(W_{\compk} | \queries) - H(W_{\compk} | A_{[1:\nNode]}^{[k]}  , \queries) \\
& = H(W_{\compk} | \queries, W_k) - H(W_{\compk} | A_{[1:\nNode]}^{[k]}  , \queries, W_k) \\
& = I(W_{\compk} ; A_{[1:\nNode]}^{[k]}  | \queries, W_k) \\
& \geq I(W_{\compk} ; A_{\nodeset}^{[k]} | \queries, W_k) \\
& \stackrel{(a)}{=}  H(A_{\nodeset}^{[k]} | \queries, \! W_k)  -  H(A_{\nodeset}^{[k]} | \queries, \! W_{[1:\nFile]}) \\
& \qquad +  H(A_{\nodeset}^{[k]} | \queries, \! W_{[1:\nFile]}, \! \secrecy) \\
& = H(A_{\nodeset}^{[k]} | \queries, W_k) - I(\secrecy ; A_{\nodeset}^{[k]} | \queries, W_{[1:\nFile]}) \\
& \geq  H(A_{\nodeset}^{[k]} | \queries, W_k, Q_{\nodeset}^{[k]}) - H(\secrecy) \\
& \stackrel{(b)}{=}  H(A_{\nodeset}^{[k]} | Q_{\nodeset}^{[k]}) - H(\secrecy) \\
& \geq H(A_{\nodeset}^{[k]} | \queries) - H(\secrecy).
\end{align*}
Equality $(a)$ holds because the answers $A_{\nodeset}^{[k]}$ are deterministic functions of the queries $\queries$, the database $W_{[1:K]}$, and the common randomness $\secrecy$.
In the proof of the converse part above, we argued that equality $(b)$ holds.

Averaging over all $\nodeset$, and from the proof in Section~\ref{sec:pfThm2_1} above,
\begin{align*}
H(\secrecy)
& \geq \frac{1}{{N \choose \max(T,E)}} \sum_{   \substack{ \nodeset \in [1:N] \\ |\nodeset|=\max{(T,E)} }   } H(A_{\nodeset}^{[k]} | \queries) \\
& \geq \frac{\max{(T,E)}}{N} H(A_{[1:\nNode]}^{[k]} | \queries) \\
& \geq \frac{\max{(T,E)}}{N -\max{(T,E)}} H(W_k).
\end{align*}

Hence, $\rho_{\textrm{T-ESPIR}} = \frac{H(S)}{H(W_k)} \geq \frac{\max{(T,E)}}{N-\max{(T,E)}} $.

\section{Discussion}
\subsection{T-BESPIR} \label{sec:T-BESPIR}
In this section, we discuss the case where an adversary has the ability to tap in on any set $\calE$ with $E$ nodes, and can overwrite the answers of any set $\calB$ with $B$ nodes (the set $\calE$ and the set $\calB$ may intersect, the adversary does not tap in on the nodes that are in $\calB$ but not in $\calE$). We argue below that the capacity of T-BESPIR is $1- \frac{2B+\max(T,E)}{N}$, with shared common randomness at least $\frac{\max(T,E)}{N-2B-\max(T,E)}$ times the size of a file.

The capacity can be achieved by simply replacing $T$ in the scheme in Section~\ref{sec:achieve_B} by $\max(T,E)$. 
User-privacy, database-privacy and decodability are guaranteed with the same arguments as in Section~\ref{sec:achieve_B}. To see that the adversary cannot obtain any information about the database, from~\eqref{eqn:answer_n}, every $E$ answers contains linearly independent combinations of $S_1, \dots, S_{\max(T,E)}$, which are uniformly and independently chosen from $\Fq$. Therefore, from any $E$ answers, the adversary cannot cancel the $S_j$'s hence obtains no information about the database.

The converse can be proved by replacing $T$ in the proof of the converse in Section~\ref{sec:converse_B} by $\max(T,E)$, and by replacing the node set $\calT$ by any node set $\calN$ with $\max(T,E)$ nodes, the same as in Section~\ref{sec:converse_E}. Because if $\max(T,E)=T$, by Lemma~\ref{thm:lemma_kkprime} and Lemma~\ref{thm:lemma_noWk}, we have that $H(A_{\nodeset}^{[k]} | W_k, Q_{\nodeset}^{[k]}) = H( A_{\nodeset}^{[k]} |  Q_{\nodeset}^{[k]} )$ holds; if $\max(T,E)=E$, from equation~\eqref{eqn:privacy_E}, $I(W_k; A_{\nodeset}^{[k]} , Q_{\nodeset}^{[k]}) = 0$, $H(A_{\nodeset}^{[k]} | W_k, Q_{\nodeset}^{[k]}) = H (A_{\nodeset}^{[k]} |  Q_{\nodeset}^{[k]} )$ also holds.
Therefore, the results are obtained by replacing $T$ in the results of Section~\ref{sec:converse_B} by $\max(T,E)$, that is, $R \leq 1- \frac{2B+\max(T,E)}{N}$ with $\rho \geq \frac{\max(T,E)}{N-2B-\max(T,E)}$.

\subsection{T-EPIR}
In this section, we discuss the case when database-privacy is not required, with colluding servers and in presence of a passive eavesdropper, hence called T-EPIR. When $E \geq T$, from the privacy of the database against the eavesdropper~\eqref{eqn:privacy_E}, $H(A_{\calE}^{[k]} | Q_{\calE}^{[k]}) = H(A_{\calE}^{[k]} | W_k, Q_{\calE}^{[k]})$. Lemma~\ref{thm:lemma_noQ} still holds for the PIR problem. With similar steps as in Section~\ref{sec:pfThm2_1}, it can be proved that $R \geq 1 -\frac{E}{N}$. The scheme in Section~\ref{sec:achieve_E} which achieves the rate of $1-\frac{E}{N}$ still works for T-EPIR problem. Hence, we can conclude that for $E \geq T$, the capacity of $T-EPIR$ equals $1-\frac{E}{N}$. The capacity of T-EPIR for the case when $E < T$ is our ongoing research.

\bibliographystyle{IEEEtran}
\bibliography{IEEEabrv,SPIR}

\end{document}